\documentstyle[twocolumn,aps]{revtex}

\begin{document}
\tolerance 50000

\draft

\twocolumn[\hsize\textwidth\columnwidth\hsize\csname @twocolumnfalse\endcsname

\title{ Theory of the optical conductivity of (TMTSF)$_2$PF$_6$ in the mid-infrared
range}

\author {Julien Favand and Fr\'ed\'eric Mila} 
\address{
      Laboratoire de Physique Quantique, Universit\'e Paul Sabatier\\
      31062 Toulouse (France)\\
      }
\maketitle

\begin{abstract}
\begin{center}
\parbox{14cm}{
We propose an explanation of the mid-infrared peak observed in the optical
conductivity of the Bechgaard salt (TMTSF)$_2$PF$_6$ in terms of electronic
excitations. It is based on a numerical calculation of the conductivity of 
the quarter-filled, dimerized Hubbard model. The main result is that, even for 
intermediate values of $U/t$ for which the charge gap is known to be
very small, the first 
peak, and at the same time the main structure, of the optical conductivity is at
an energy of the order of the dimerization gap, like in the infinite $U$ case.
This surprising effect is a consequence of the optical selection rules.
}
\end{center}
\end{abstract}

\pacs{
\hspace{1.9cm}
}
\vskip2pc]

\section{Introduction}

It is by now well established experimentally that there is a well defined
structure in the mid-infrared optical conductivity of the Bechgaard salt 
(TMTSF)$_2$PF$_6$\cite{Jacobsen,Pedron,Dressel}. According to the most recent
data of Dressel et al\cite{Dressel}, this structure moves from about 
1000 cm$^{-1}$ at room temperature to 200 cm$^{-1}$ at 20 K, while its intensity
increases upon lowering the temperature. Given the width and the intensity of
this structure, it seems difficult to explain it in terms of phonons,
and the most natural thing to do it to look for an
explanation in terms of electronic excitations. 

It is not so easy however to understand why there should be
an absorption due to electronic transitions in
that energy range. To see that, let us consider the simplest description of the
Bechgaard salt that contains the essential physics, namely the quarter-filled,
dimerized 
Hubbard model described by the following Hamiltonian:
\begin{eqnarray}
H=-t_1\sum_{i\ {\rm even},\sigma}(c{}^\dagger_{i+1\sigma}c{}^{\ }_{i\sigma}+h.c.) \cr
-t_2\sum_{i\ {\rm  odd},\sigma}(c{}^\dagger_{i+1\sigma}c{}^{\ }_{i\sigma}+h.c.)
+U\sum_i n_{i\uparrow}n_{i\downarrow} 
\end{eqnarray}

where $c{}^\dagger_{i\sigma},c{}^{\ }_{i\sigma}$ create and annihilate holes in
the HOMO of the TMTSF molecules.
For (TMTSF)$_2$PF$_6$, reasonable parameters  are $t_1$=250 meV, $t_2/t_1=.9$
and $U/t_1=5$. 
Note that a Luttinger liquid description would not be of much use 
here given the energy
range we are interested in.

Now, let us see which electronic transitions could be present in such a model.
If the interactions are ignored, 
the only allowed transitions are the vertical transitions from the lower band to
the upper band. They give rise to a continuum starting at  
$\Delta_0 \equiv 2\sqrt{t_1^2+t_2^2} \simeq  
670$ meV, which is 
roughly
one order of magnitude too large. Including $U$ and assuming that $U/t_1=5$  
will induce transitions at
energy $U\simeq 1.25$ eV, which is even larger. 
In fact there is only
one characteristic energy scale that has the right order of magnitude, namely
the dimerization gap $ \Delta_D=2(t_1-t_2)=50$ meV.
This led Pedron et al \cite{Pedron} to assume that the double
occupancy of a site must be excluded, or equivalently that $U$ is infinite.
Then the charge carriers can be described as a half-filled system of 
spinless fermions, in which case the vertical transition has an energy of
$\Delta_D$.
There is a problem however with that explanation. If $U$ is infinite, then we
know that the charge gap $\Delta_{\rho}$ is equal to the dimerization gap 
$\Delta_D$\cite{Penc}.
But such a large charge gap is inconsistent with the experimental fact that the
conductivity remains metallic down to very low temperatures. In fact, the
metallic character of the conductivity puts an upper bound on the charge gap
which is consistent with the above-mentioned parameters\cite{Penc}.

In this paper, we show that the structure observed in the optical conductivity
is actually consistent with the 
model of Eq. (1) with $t_1$=250 meV, $t_2/t_1=.9$
and $U/t_1=5$.
The main point
is that the charge excitations with energy around $\Delta_{\rho}$
responsible for the metallic 
conductivity cannot be seen in the optical conductivity because of the optical
selection rules. The first
allowed transition occurs at an energy 
of the order of the dimerization gap $\Delta_D$, which is much larger than  
$\Delta_{\rho}$ for these values of the parameters.

The paper is organized as follows. In section II, we first discuss the general 
properties 
of the optical conductivity in the model of Eq. (1). In section III,
we describe the numerical method and we derive the form of the finite-size 
corrections we can expect for the position of the first peak of the optical
conductivity. In sections IV and V, we present the results for large and small
dimerization respectively. Finally, a discussion of the results in connection 
to (TMTSF)$_2$PF$_6$ is given in section VI.

\section{General properties of the conductivity}

Let us start with a qualitative description of the optical conductivity in
various cases of the Hamiltonian of Eq. (1) at quarter-filling. They are
illustrated in Fig. \ref{fig1}.

{\it a) Non interacting electrons on a non-dimerized lattice ($U=0$,
$t_2=t_1$):} In that trivial limit, the only contribution is of course the zero
frequency Drude peak.

{\it b) Interacting electrons on a non-dimerized lattice ($U\ne 0$,
$t_2=t_1$):} There is still a Drude peak because the system remains metallic, but
there is also some incoherent spectral weight around $\omega = U$.

{\it c) Non interacting electrons on a dimerized lattice ($U=0$,
$t_2\ne t_1$):} The dimerization splits the dispersion into 2 bands with 
the equation:    
\begin{equation} 
E(k) = \pm \sqrt{t_1^2+t_2^2+2t_1t_2\cos(2ka)} 
\end{equation}
where $2a$ is the lattice parameter.
The conductivity exhibits a Drude peak since the system is metallic, 
and a continuum between $\Delta_0$ and $2(t_1+t_2)$ due to vertical interband transitions. See Fig. \ref{fig2}. 

{\it d) Interacting electrons on a dimerized lattice ($U\ne 0$,
$t_2\ne t_1$):} In that case the repulsion opens a gap because the lower band is
effectively half-filled. 
So the system becomes insulating and the Drude peak disappears.
Now the incoherent and Drude weight are related by the sum-rule
\begin{equation}
\int_{0}^{\infty}\sigma(\omega)d\omega = \frac{\pi e^2}{2L}<-T> 
\end{equation}
where $<T>/L$ is the expectation value of the kinetic energy per-site in the ground state.
This quantity is not dramatically reduced by the interaction\cite{Mila}, and 
the lost  Drude weight must be redistributed as an
incoherent, low energy background. Part of the weight can also migrate to the
upper Hubbard band at an energy of order $U$.

If $U$ is very large, the low energy, incoherent weight must be located around
$\Delta_D$ to become the zone-boundary interband transition of spinless fermions
in the $U=+\infty$ limit described by Pedron et al\cite{Pedron}. In the
following our main goal is to determine whether for realistic parameters, i.e.
for intermediate values of $U/t_1$, there is still a dominant low energy
structure in the conductivity, and where it is located.

\section{The numerical method}

Our aim is to calculate the optical conductivity of the model at zero 
temperature. 
Starting from its definition as the current-current correlation 
function, the real part of the conductivity can also be written\cite{Dagotto}:
$$
\sigma(\omega)=D\delta(\omega) 
+\frac{\pi e^2}{L}\sum_{n \neq 0}
\frac{|<\psi_0|\hat{\j }|\psi_n>|^2}{E_n-E_0}\delta(\omega-E_n+E_0)
$$
In this formula, $E_0$ is the ground state energy, $n$ labels the excited 
states and $\hat{\j }$ is the paramagnetic current operator.
The weight of the Drude peak $D$ is calculated from the Kohn relation
\cite{Kohn,Zotos,Shastry}:
\begin{equation}
D=\frac{\pi}{L} \frac{\partial^2 E_0}{\partial \phi^2}\huge|_{\phi=0}
\end{equation}
where $E_0(\phi)$ is the ground state energy as a function of the twist in 
the boundary conditions, while the incoherent part can be obtained as a continued
fraction using Lanczos algorithm\cite{Gagli}. Extensive calculations of that
sort have been performed for 2D models in the context of high T$_c$
superconductors\cite{dagotto2}.

We have performed exact diagonalizations on finite size clusters with
$ L= 8,12,16,20 $ sites. To reach the
thermodynamic limit ($L=+\infty$) we have tried to perform a finite-size scaling
of the results. 
The finite size corrections turn out to be very large concerning 
the location $\Omega$ of the first peak
in the conductivity for the following reasons:
In the infinite U limit, the charge and spin variables are decoupled
(see Ogata and Shiba \cite{Ogata}). The
charge part is that of spinless fermions with twisted boundary conditions
and this twist is given by the momentum of the spin part.
Now, for $U=+\infty$ there is no energy associated to spin excitations, and the
spinless fermions are free to choose the boundary conditions that minimize this
energy, namely 
antiperiodic boundary conditions. In that case, 
the allowed values of the momentum are given by $k=(2\nu+1)\pi/La$ 
(see Fig. \ref{fig2}). The zone
boundary $k=\pi/2a$ does not belong to the allowed momenta 
and the larger occupied $k$ is at $\pi/2a-\pi/La$.
So the first peak in the conductivity is located at
\begin{equation} 
\Omega = \Delta_D(L) = \sqrt{t_1^2+t_2^2+2t_1t_2\cos(\pi-2\pi/L)} 
\end{equation}
In other words, there is a finite--size correction to the known result 
$\Omega = \Delta_D$
in the
thermodynamic limit.
In the case of large but finite U, the spin energy is still negligible compared 
to that of the charge degrees of freedom, 
and the charge part keeps the same boundary conditions. 
This suggests that the finite--size corrections to
$\Omega^2$ will scale according to $\cos(\pi-2\pi/L)$. We shall see that this
scaling form is actually remarkably accurate down to rather small values of
$U/t_1$.

\section{Large dimerization}

For clarity, let us start with a case where all the important features of the conductivity
can be best seen in spite of the limitations due to the finite size of the 
clusters. In units
of $t_1$ we choose $t_2=0.5$ and study various repulsions.
On Fig. \ref{fig3} we show the influence of the cluster size 
on the incoherent spectum for a given repulsion.
This incoherent part exhibits a sharp peak which is shifted toward low
frequencies
when the size increases, while its relative weight is more or less constant. 
Figure \ref{fig4} shows the effect of the repulsion 
($U/t_1$ change from 0 to 10) for L=16.
The main peak increases with size and approaches  $\Delta_D(L)$ for large $U$
as expected. 

Because of the dimerization, the lower band is half filled and the repulsion induces an
insulating behavior in the thermodynamic limit. So the Drude weight 
given by Eq. (5), which does not vanish for finite systems, should scale to zero
upon increasing the size of the cluster. Our numerical results are effectively
consistent
with a vanishing Drude weight in the thermodynamic limit, although the finite
size scaling function is not very clear (see Fig \ref{fig5}).

We have also checked the sum-rule of Eq. (3).
The finite range of integration could bring problems
in practical computations, but in the case of
quarter--filled systems, there is essentially no weight at high
frequencies\cite{Maldague,Eskes}
and one can safely stop the integration at $\omega=2U$.
In all cases we found that the sum--rule was satisfied with an accuracy better
than 1\%.

In order to show that the incoherent conductivity exhibits a well defined
structure at low energy, we have calculated the relative weight of
the first peak with respect to the total incoherent part.
The results are given in  Table \ref{tableau2}. This proportion is insensitive 
to the size
and increases with the repulsion.
This suggests that the first peak will dominate the incoherent part 
in the thermodynamic limit,
even for intermediate repulsions.

To find the location $\Omega$ of this peak, we have tried several finite--size
scaling. The only way to obtain a good scaling is to plot $\Omega^2$ versus
$\cos(\pi-2\pi/L)$. Some results are presented in Fig. \ref{fig6}. The extrapolated
values are given in Table II. In the present case, the location of the first
peak roughly follows the charge gap $\Delta_\rho$, which is relatively 
large for this value of the dimerization.

\section{Small dimerization}

We now turn to a smaller dimerization, namely  $t_2=0.9t_1$, having in mind
intermediate repulsions $U\simeq 5t_1$.
The incoherent conductivity
$\sigma_{inc}$ calculated for L=20 is shown in figure \ref{fig7}.  As in the
previous case $\sigma_{inc}$ is still dominated by
 its first peak. The relative importance of this peak increases
 from 34.3\% on 8 sites to
 53.1\% on 20 sites.

In the present case $\sigma_{inc}$ represents a rather small part of 
the oscillator 
strength
(from 2 \% for 8 sites to 4.5 \% for 20 sites). Because of the weak 
dimerization, the charge
gap is small ($\Delta_{\rho}=0.02t_1$), and the corresponding length
$v_F/\Delta_\rho$ is much larger than the sizes we can reach.
So the Drude peak takes almost all the oscillator strength in our simulation. 
But this peak will certainly 
disappear in the thermodynamic limit since the system is insulating. So what
really matters is whether the relative weight of the first peak be sizable, which
it is.

Concerning the location of the first peak, we found that 
scaling $\Omega^2$ with $\cos (\pi-2\pi/L)$ was still very accurate and 
for $U=5t_1$ this scaling leads
$\Omega/t_1=0.17$ in the thermodynamic limit. Now, for $t_2=.9t_1$
and $U=5t_1$, the charge gap $\Delta_\rho$ equals 0.02 and the dimerization gap 
$\Delta_D$ equals 0.2. So we found that the first peak appears at an energy
which is of the order of the magnitude of the dimerization gap, and that there
is no weight at energies corresponding to the charge gap. How can we understand
this result?
If the charge gap is identified to the lowest excitation of the system which
leaves the total
spin S unchanged, the difference between $\Omega$ and $\Delta_{\rho}$ should
be directly observable on
the excitation spectrum. Such a spectrum is plotted in Fig. \ref{fig8}. The lowest charge
excitation occurs at the border of the Brillouin zone, while $\Omega(L)$ corresponds to a vertical
excitation keeping S=0. 
Besides, the first singlet state coupled to the groundstate by the current
operator is not the lowest S=0 excited state but the third one\cite{note}.
This is presumably due to the fact that the current operator being odd under the
inversion, only odd states can be coupled to the groundstate, which is
even. 
$\Omega(L)$ and $\Delta_\rho$ 
are thus clearly different for L finite. Besides, they have 
very different scalings: On one hand
$\Delta_{\rho}$  goes to a very small value like $1/L$ as shown 
on Fig \ref{fig9}.
The exact value 0.02  provided by a perturbative calculation can actually not be
identified by such a scaling.
On the other hand, $\Omega(L)^2$ scales as $\cos(\pi-2\pi/L)$ toward 0.17.

So our results indicate that the first peak of the optical conductivity  is
located at an energy much larger than the charge gap. We cannot exclude that,
upon increasing the size, other states will appear that have the right symmetry
to be coupled to the groundstate by the current operator. Such states could for
instance involve several excitations of momentum $\pi$ and of energy of order
$\Delta_\rho$ leading to a small threshold at $2\Delta_\rho$ in the optical
conductivity. But according to the increase with $L$ of the relative weight of
the first peak (from 34.3\% for 8 sites to 53.1\% for 20 sites), the feature at
$\Omega\simeq0.17t_1$ should in any case remain the prominent structure of the
conductivity in the thermodynamic limit. Finally, the difference with the case
treated in the previous section presumably comes from the fact that the
excitations have very little dispersion when the dimerization is large because
the bands are flat.

So it seems that the explanation proposed by Pedron et al\cite{Pedron} is
essentially valid: For intermediate values of $U/t_1$ and small dimerization,
the conductivity has a large peak around the dimerization gap $\Delta_D$
although the charge gap is very small.

\section{Discussion}

In conclusion, we have shown that the model of Eq. (1) with reasonable
parameters leads to a peak in the optical conductivity at an energy of about
$0.17t_1 \simeq 40$
meV. This energy has the right order of magnitude. So this provides a good
candidate to explain the mid-infrared structure observed in the Bechgaard salt 
(TMTSF)$_2$PF$_6$. What about the fine details of this mid-infrared structure?
As stated in the Introduction, this peak moves toward lower energies when the
temperature is decreased, and its weight increases. Both features are actually
quite natural consequences of the present explanation.

Let us first consider the fact that the peak moves to lower energies when the
temperature is lowered. The Hamiltonian of Eq. (1) , which is purely 1D, is not
an accurate description of the electronic structure of the Bechgaard salts at
low temperatures where 2D effects start to dominate. Looking at the actual 2D
band structure calculated on the basis of the structure deduced from X-Ray
measurements performed at 300 K and 4 K, one can see that there is a clear
evolution leading to a much smaller dimerization gap at low temperatures.
Whether this gap still yields a structure in the optical conductivity when the
system must be considered as 2D cannot be concluded on the basis of the present
calculation and should be checked independently, but it seems likely that the
structure we have calculated will move smoothly to lower energies.

Concerning the intensity, the data of Dressel et al\cite{Dressel} suggest that 
the increase of weight of that structure upon lowering the temperature
is accompanied by a decrease of the
weight of the background, which can be interpreted as a broadened Drude peak.
Although the model we are looking at is insulating at zero temperature, we
expect a Drude peak to be restored at temperatures comparable to the charge gap 
$\Delta_\rho \simeq .02 t_1 \simeq 40$ K. It seems plausible that the weight of
this peak will thus increase with temperature in the range 50-100 K. Note that
the fact that the system is metallic below 40K is not incompatible with 
$\Delta_\rho \simeq 40$ K because the physics is certainly 2D below that
temperature\cite{Wzietek}, and the 1D description does not apply anymore.
It would be interesting to check if this evolution can be confirmed by 
Quantum Monte
Carlo simulations.

We acknowledge very useful discussions with L. Degiorgi, M. Dressel and X.
Zotos. We are especially grateful to D. Poilblanc for very useful comments on
several aspects of the calculation of the conductivity by exact diagonalization
of finite clusters. 
The numerical calculations have been performed on the C-94 and C-98 of
IDRIS (Orsay).

\begin{table} [h]
\caption{Relative weight of the first peak in the incoherent part,
 with boundary conditions ensuring a
 non-degenerate ground state. \label{tableau2}}
\end{table}

\begin{table} [h]
\caption{Comparison between the charge gap and the location of the first peak in the
 optical conductivity
 for large dimerization ($t_2=0.5t_1$) and various repulsions. The value of
 $\Delta_\rho$ are taken from Ref. 8. \label{tableau3}}
\end{table}

\begin{figure}[ht]
\caption{Schematic picture showing the influence of the repulsion and of the
dimerization on the structure of the conductivity. $\Delta_D=2(t_1-t_2)$ 
and  $\Delta_0=2\sqrt{t_1^2+t_2^2}$. Fig. (d) is typical of large values
 of $U$. For clarity, the $\delta$ peaks of the
incoherent part of the conductivity have been broadened.}
\label{fig1}
\end{figure}

\begin{figure}[ht]
\caption{Dispersion in the non-interacting case with dimerization. (1) Vertical transition in the
non interacting case at $\omega=\Delta_0$. (2) Finite size dimerization gap $\Delta_D(L=16)$.
 (3) Dimerization gap.  The circles (resp. triangles) are the allowed wave vectors for L=16
 and periodic (resp. antiperiodic) boundary conditions.} 
\label{fig2}
\end{figure}

\begin{figure}[ht]
\caption{Evolution of the optical conductivity with the size of the cluster,
 for $t_2=0.5t_1$, $U=5t_1$ and $\epsilon=0.05t_1$
 (width of the $\delta$ function).}
\label{fig3}
\end{figure}

\begin{figure}[ht]
\caption{Evolution of the optical conductivity with the on site repulsion U
 for $t_2=0.5t_1$, $L=16$ and $\epsilon=0.05t_1$
 (width of the $\delta$ function).}
\label{fig4}
\end{figure}

\begin{figure}[ht]
\caption{Finite size scaling of the relative weight of the Drude peak
($D(L)/(\pi e^2<-T>/L)$) for $t_2=0.5t_1$.}
\label{fig5}
\end{figure}

\begin{figure}[ht]
\caption{Finite size scaling of the square of the frequency $\Omega(L)$
 (in units of $t_1$) of the first peak of the incoherent conductivity.}
\label{fig6}
\end{figure}

\begin{figure}[ht]
\caption{The optical conductivity for $t_2=0.9t_1$, $U=5t_1$, $L=20$
 and $\epsilon=0.05t_1$ (width of the $\delta$ function).}
\label{fig7}
\end{figure}

\begin{figure}[ht]
\caption{The first excitations versus the impulsion for $t_2=0.9$, $U=5$, $L=20$.
 The impulsion $k$ is related to $\nu$ by $k=-\pi/L+2\pi\nu/L$ because of antiperiodic
 boundary conditions. The circles (resp. triangles) represent states
 with a total spin S=0 (resp. $S \neq 0$). (1) First
excitation coupled to the ground state by the current operator. (2) First singlet excitation.}
\label{fig8}
\end{figure}

\begin{figure}[ht]
\caption{Finite size scaling of the lowest charge excitation for 
 $t_2=0.9$ and $U=5$.}
\label{fig9}
\end{figure}

\begin{table} [h]
\begin{center} 
\begin{tabular}{lccc} \hline
  &  $U=2.5$ & $U=5$ & $U=10$ \\
\hline
$L=8$ & 46.8\% & 69.3\% & 75.2\% \\
\hline
$L=12$ & 49.6\% & 67.3\% & 74.6\% \\
\hline
$L=16$ & 51.5\% & 64.3\% & 68.2\% \\
\hline
\end{tabular}
\end{center}
%\label{tableau}
\end{table}

\begin{table} [h]
\begin{center} 
\begin{tabular}{lccc} \hline
  &  $U=2.5$ & $U=5$ & $U=10$ \\
\hline
$\Omega$ & $0.22t_1$ & $0.50t_1$& $0.74t_1$ \\
\hline
$\Delta_{\rho}$ & $0.2t_1$  & $0.45t_1$ & $0.7t_1$ \\
\hline
\end{tabular}
\end{center}
%\label{tableau}
\end{table}

\onecolumn
\newpage
%
%Fig. 1
%
\begin{figure}[ht]
\vbox to 283bp {%\vfil
\centerline{\hbox to 407bp {\includegraphics{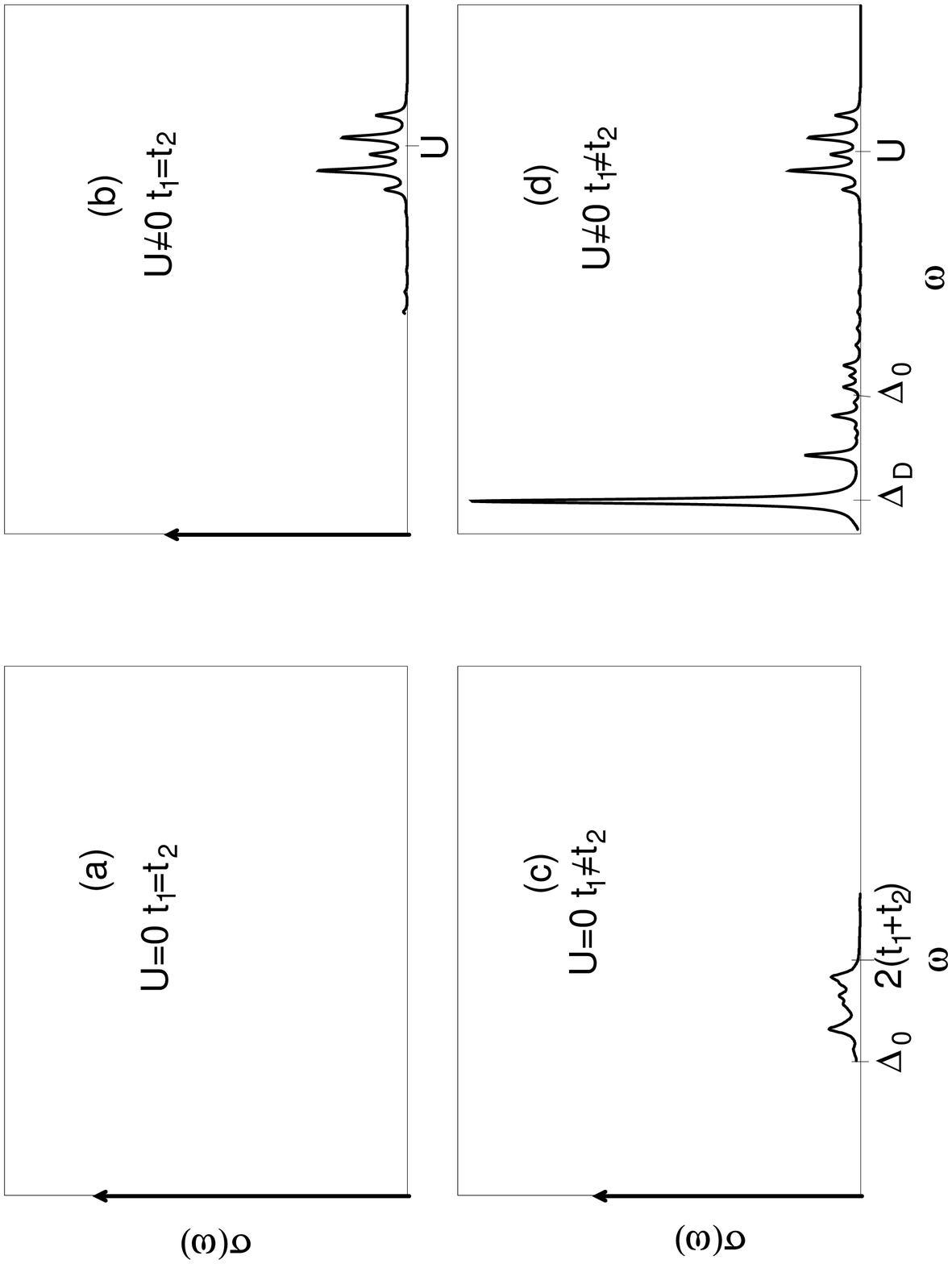}\hfil}}\vfil}
\end{figure}

%
%Fig. 2
%
\begin{figure}[ht]
\vbox to 283bp {%\vfil
\centerline{\hbox to 407bp {\includegraphics{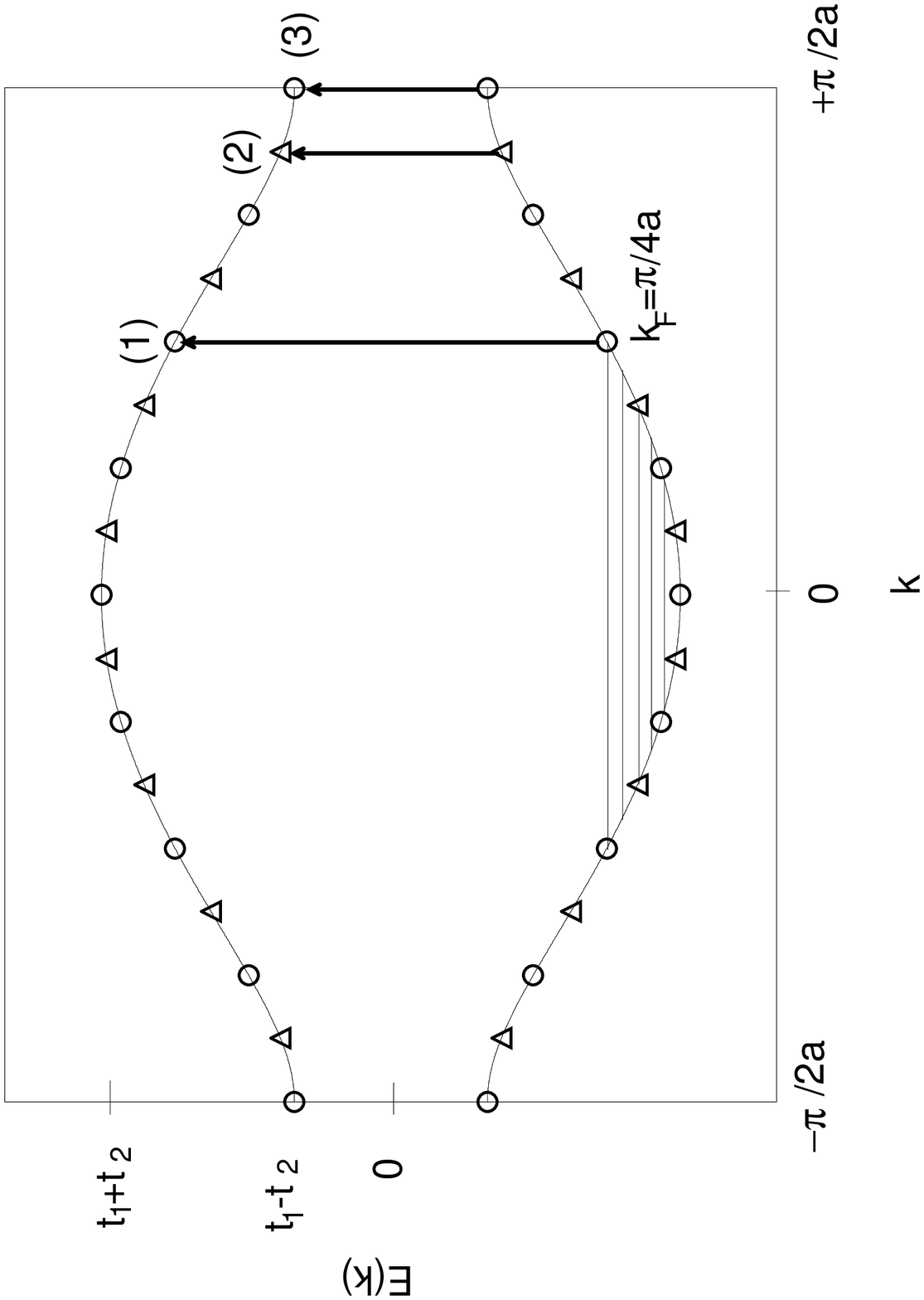}\hfil}}\vfil}
\end{figure}

\newpage
%
%Fig. 3
%
\begin{figure}[ht]
\vbox to 283bp {%\vfil
\centerline{\hbox to 407bp {\includegraphics{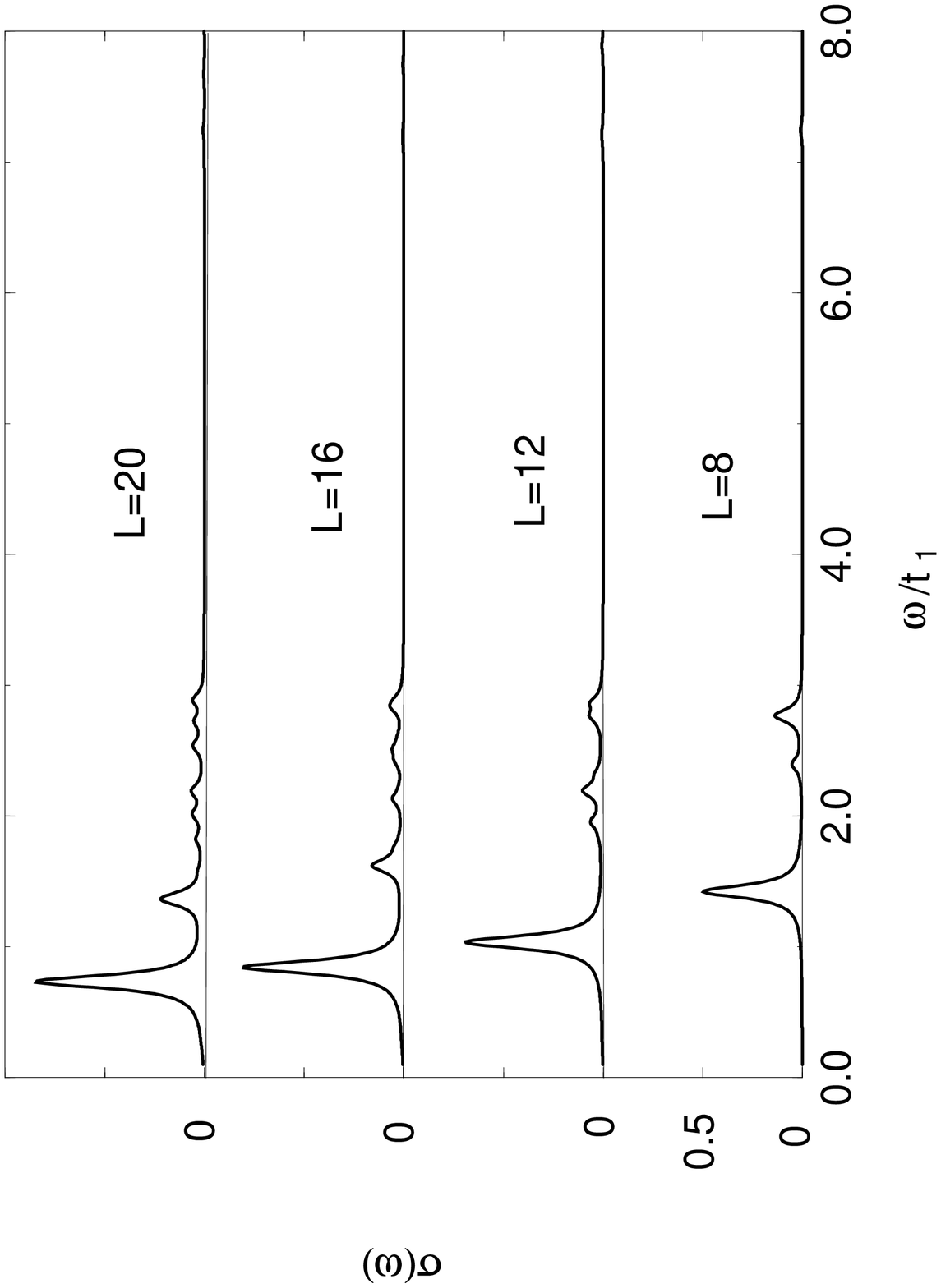}\hfil}}\vfil}
\end{figure}

%
%Fig. 4
%
\begin{figure}[ht]
\vbox to 283bp {%\vfil
\centerline{\hbox to 407bp {\includegraphics{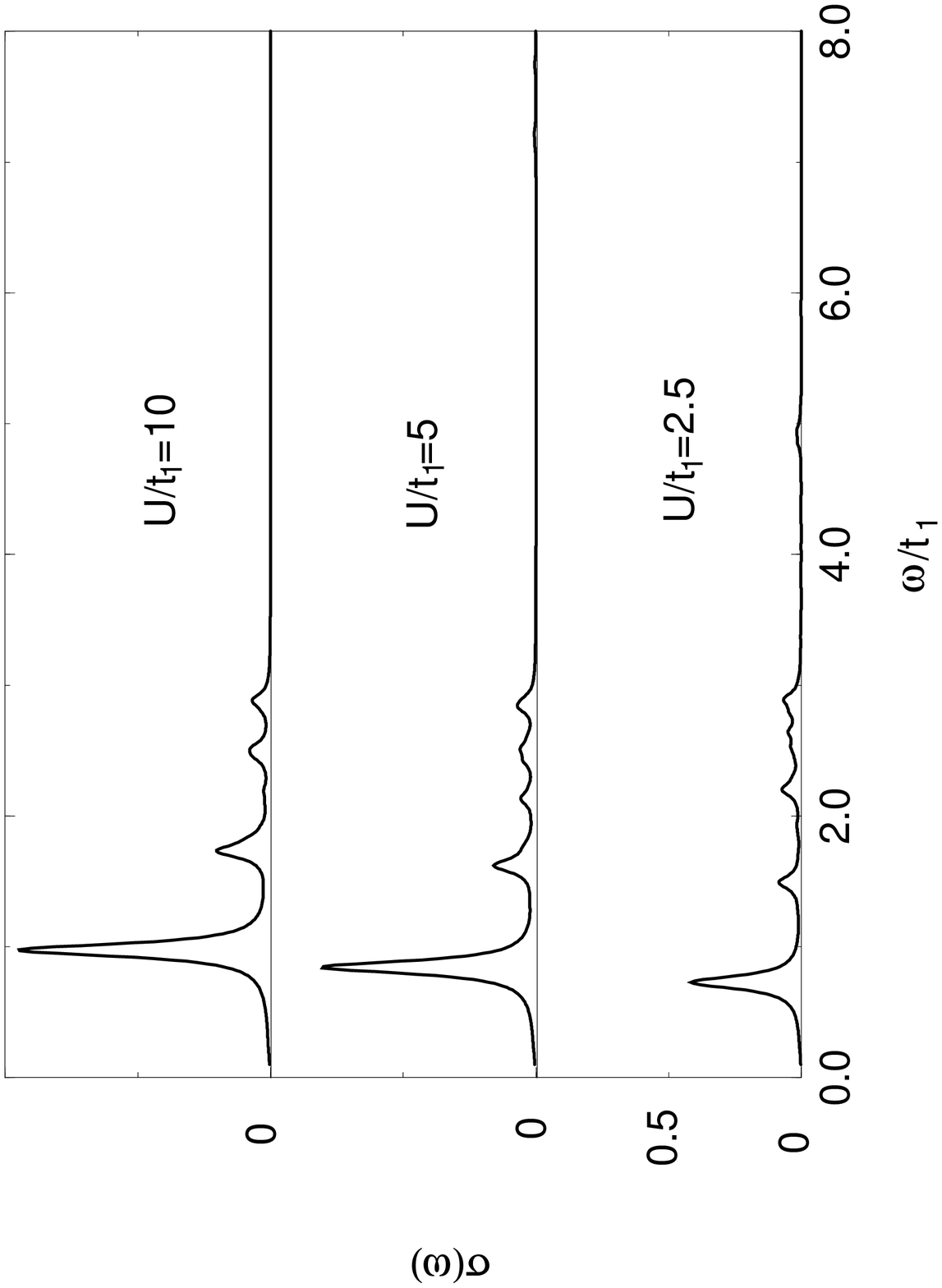}\hfil}}\vfil}
\end{figure}

\newpage
%
%Fig. 5
%
\begin{figure}[ht]
\vbox to 283bp {%\vfil
\centerline{\hbox to 407bp {\includegraphics{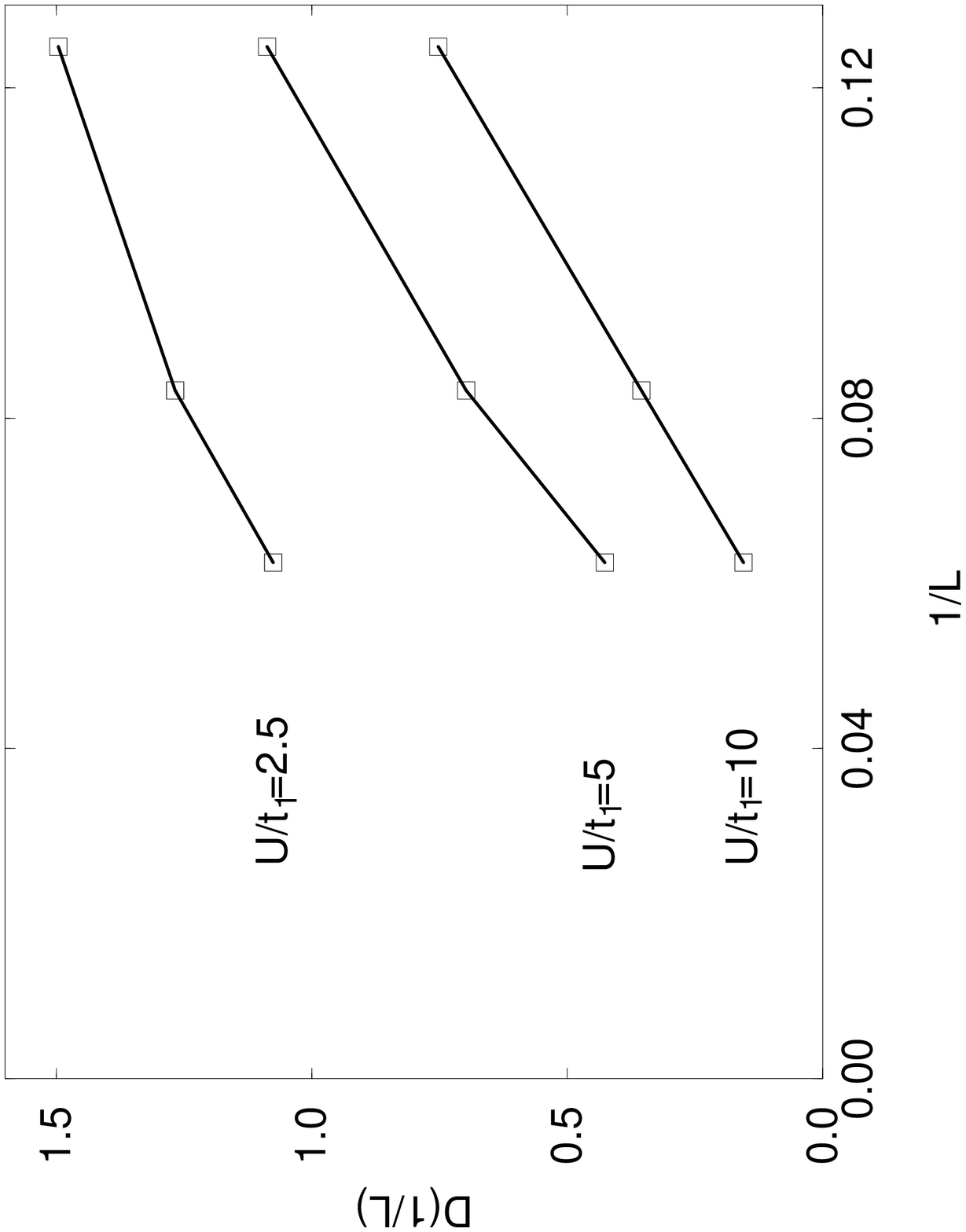}\hfil}}\vfil}
\end{figure}

%
%Fig. 6
%
\begin{figure}[ht]
\vbox to 283bp {%\vfil
\centerline{\hbox to 407bp {\includegraphics{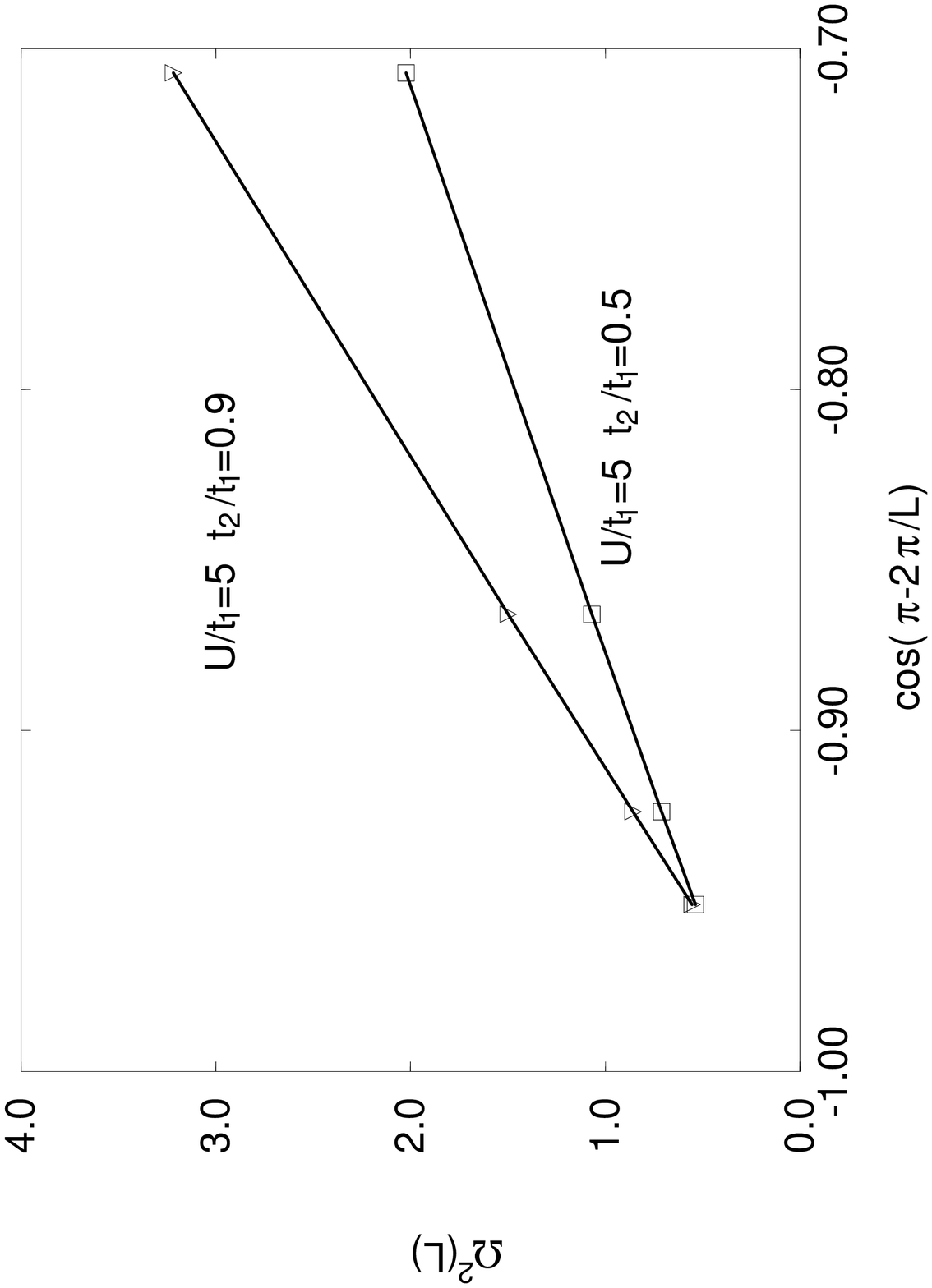}\hfil}}\vfil}
\end{figure}

\newpage
%
%Fig. 7
%
\begin{figure}[ht]
\vbox to 283bp {%\vfil
\centerline{\hbox to 407bp {\includegraphics{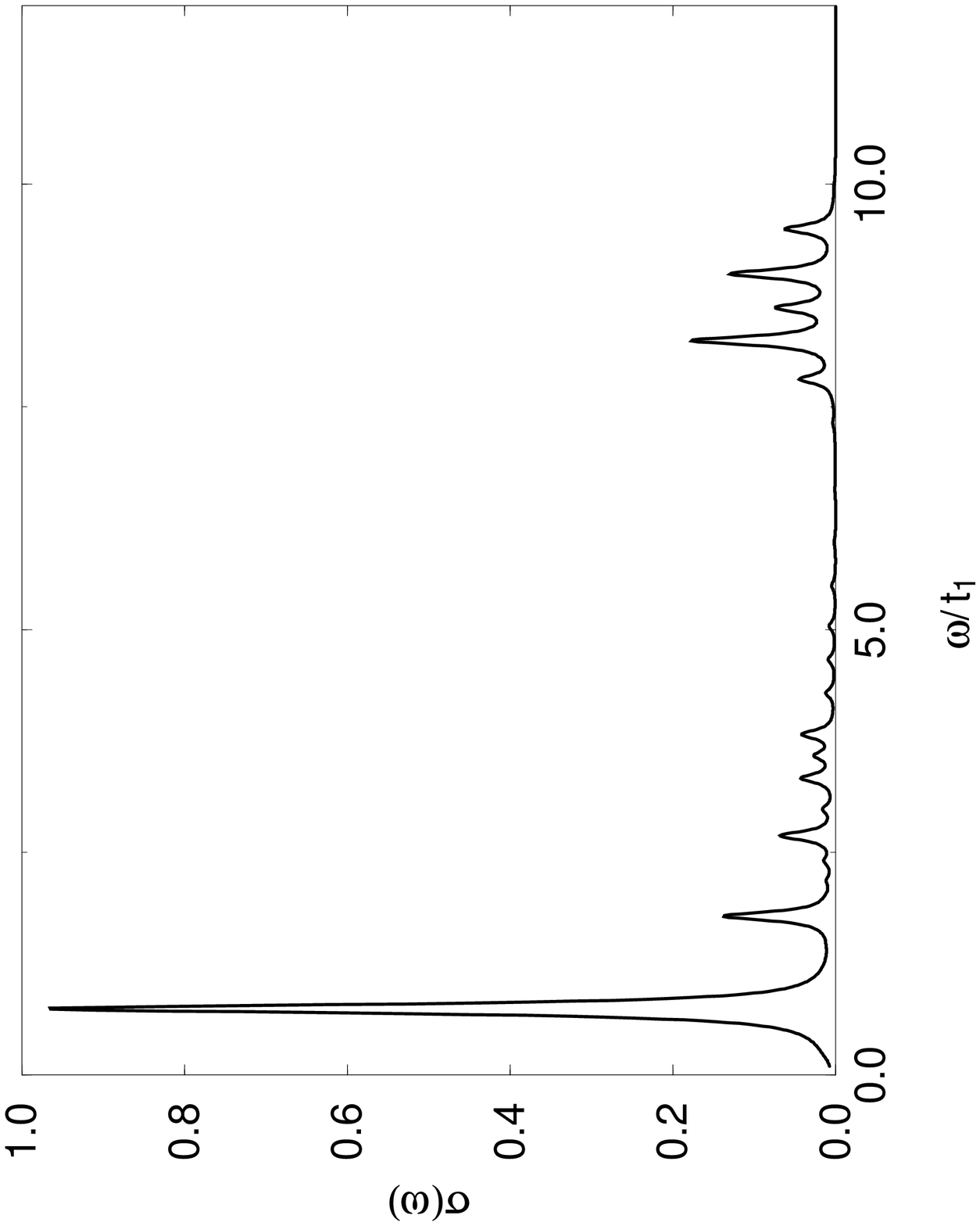}\hfil}}\vfil}
\end{figure}

%
%Fig. 8
%
\begin{figure}[ht]
\vbox to 283bp {%\vfil
\centerline{\hbox to 407bp {\includegraphics{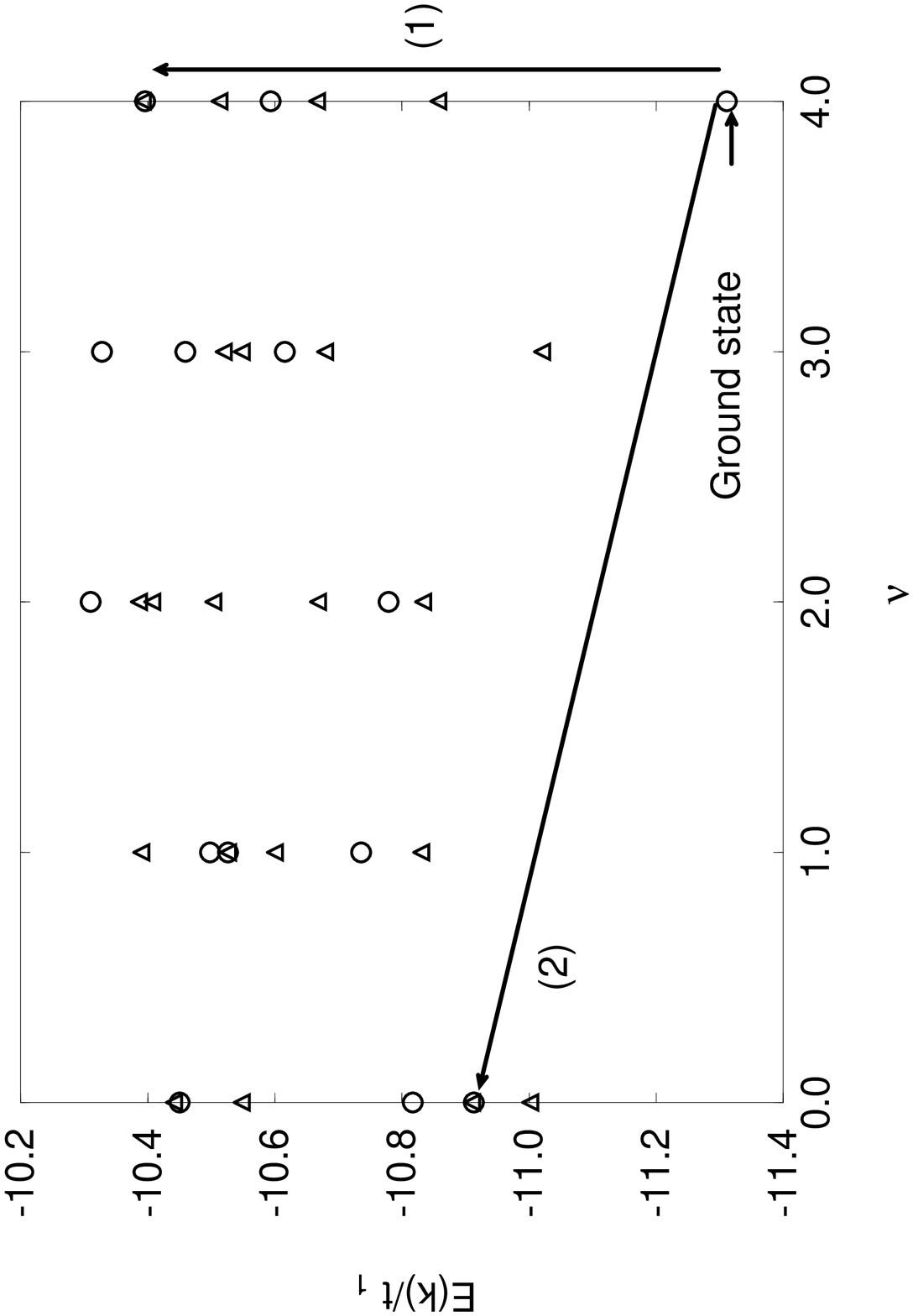}\hfil}}\vfil}
\end{figure}

\newpage
%
%Fig. 9
%
\begin{figure}[ht]
\vbox to 283bp {%\vfil
\centerline{\hbox to 407bp {\includegraphics{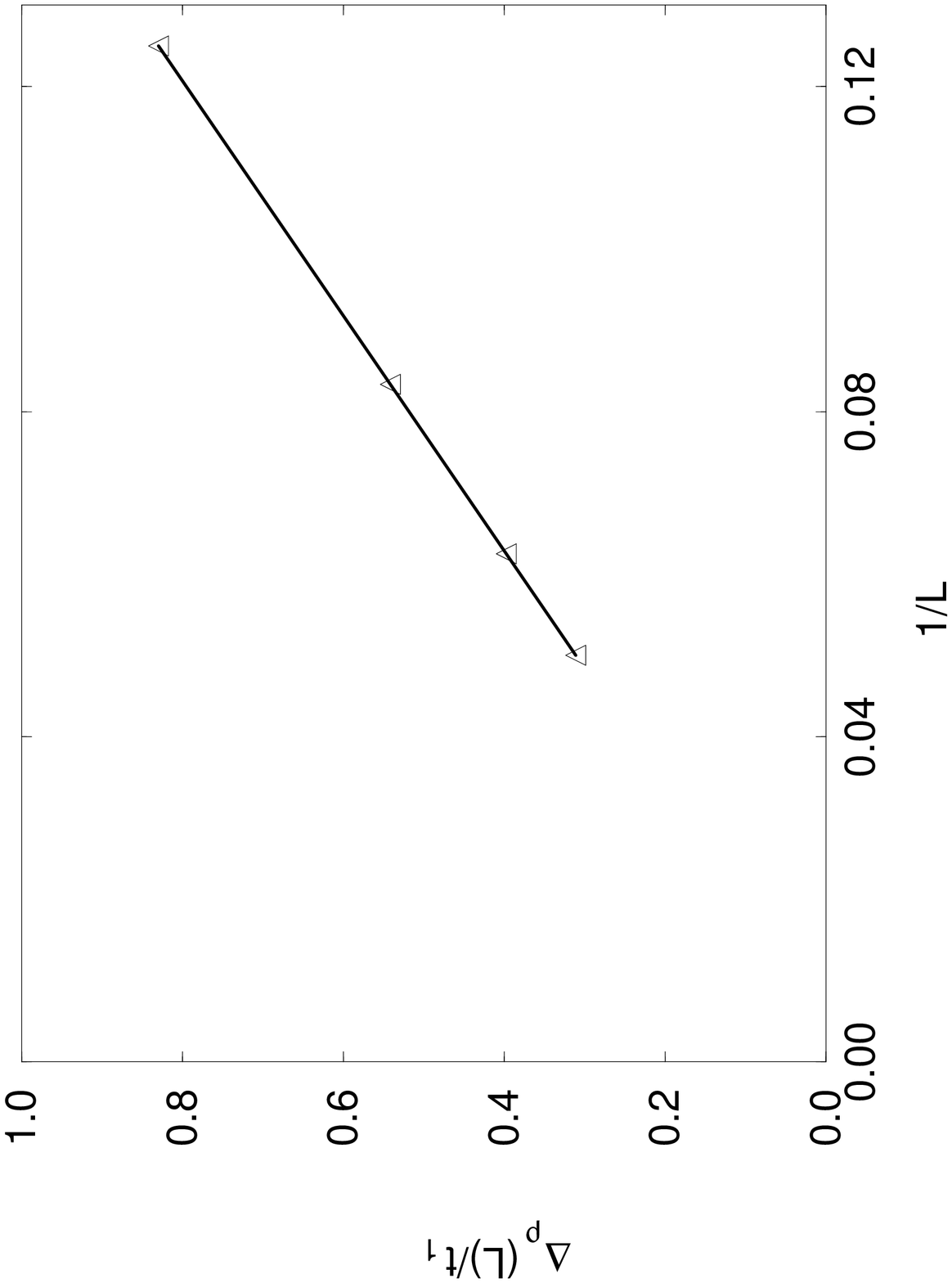}\hfil}}\vfil}
\end{figure}


\begin{thebibliography}{99}

\bibitem{Jacobsen} C. S. Jacobsen, D. B. Tanner and K. Bechgaard, Phys. Rev. B
{\bf 38}, 7019 (1983).

\bibitem{Pedron} D. Pedron, R. Bozio, M. Meneghetti, C. Pecile, Phys. Rev. B {\bf
49}, 10893 (1994)

\bibitem{Dressel} M. Dressel, A. Schwartz, G. Gruner, L Degiorgi, unpublished.

\bibitem{Dagotto} See e.g. E. Dagotto, Rev. Mod. Phys. {\bf 66}, 763 (1994) and references therein.

\bibitem{Gagli} E.R. Gagliano and C.A. Balseiro, Phys. Rev. B {\bf 38}, 11766 
(1988).

\bibitem{dagotto2} See e.g. E. Dagotto, A. Moreo, F. Ortolani, D. Poilblanc, 
J. Riera,
Phys. Rev. B {\bf 45}, 10741 (1992), and references therein.

\bibitem{Mila} F. Mila, Phys. Rev. B {\bf 52}, 4788 (1995).

\bibitem{Penc} K. Penc and F. Mila, Phys. Rev.B {\bf 50}, 11429 (1994).

\bibitem{Kohn} W. Kohn, Phys. Rev. {\bf 133}, A171 (1964).

\bibitem{Zotos} X. Zotos, P. Prelovsek, I. Sega, Phys. Rev. B {\bf 42}, 8445
(1990).

\bibitem{Shastry} B. S. Shastry and B. Sutherland, Phys. Rev. Lett.{\bf 65}, 243
(1990).


\bibitem{Ogata} M. Ogata and H. Shiba, Phys. Rev. B {\bf 43}, 8401 (1990).

\bibitem{Maldague} P. F. Maldague, Phys. Rev. B {\bf 16}, 2437 (1977).

\bibitem{Eskes} H. Eskes and A. M. Oles, Phys. Rev. Lett. {\bf 73}, 732 (1994).

\bibitem{Wzietek} P. Wzietek, F. Creuzet, C. Bourbonnais, D. J\'erome, L
Bechgaard, P. Batail, J. Physique I {\bf 3}, 171 (1993).

\bibitem{note} On Fig. 8, the second S=0 excited state cannot be seen because it
is almost degenerate with the third one.

\end{thebibliography}
\end{document}